\def \SAIT #1 #2 {{\em Mem.\ Soc.\ Astron.\ It.\/} {\bf #1}, #2}
\def \MESS #1 #2 {{\em The Messenger\/} {\bf #1}, #2}
\def \ASTRNACH #1 #2 {{\em Astron. Nach.\/} {\bf #1}, #2}
\def \AAP #1 #2 {{\em Astron. Astrophys.\/} {\bf #1}, #2}
\def \AAPs {{\em Astron. Astrophys.\/}} 
\def \AAL #1 #2 {{\em Astron. Astrophys. Lett.\/} {\bf #1}, L#2}
\def \AAR #1 #2 {{\em Astron. Astrophys. Rev.\/} {\bf #1}, #2}
\def \AAS #1 #2 {{\em Astron. Astrophys. Suppl. Ser.\/} {\bf #1}, #2}
\def \AJ #1 #2 {{\em Astron. J.\/} {\bf #1}, #2}
\def \ANNREV #1 #2 {{\em Ann. Rev. Astron. Astrophys.\/} {\bf #1}, #2}
\def \APJ #1 #2 {{\em Astrophys. J.\/} {\bf #1}, #2}
\def \APJL #1 #2 {{\em Astrophys. J. Lett.\/} {\bf #1}, L#2}
\def \APJS #1 #2 {{\em Astrophys. J. Suppl.\/} {\bf #1}, #2}
\def \APSS #1 #2 {{\em Astrophys. Space Sci.\/} {\bf #1}, #2}
\def \ASR #1 #2 {{\em Adv. Space Res.\/} {\bf #1}, #2}
\def \BAIC #1 #2 {{\em Bull. Astron. Inst. Czechosl.\/} {\bf #1}, #2}
\def \JSQRT #1 #2 {{\em J. Quant. Spectrosc. Radiat. Transfer\/} {\bf #1}, #2}
\def \MN #1 #2 {{\em Mon. Not. R. Astr. Soc.\/} {\bf #1}, #2}
\def \MEM #1 #2 {{\em Mem. R. Astr. Soc.\/} {\bf #1}, #2}
\def \PLR #1 #2 {{\em Phys. Lett. Rev.\/} {\bf #1}, #2}
\def \PR #1 #2 {{\em Phys. Rev.\/} {\bf #1}, #2}
\def \PASJ #1 #2 {{\em Publ. Astron. Soc. Japan\/} {\bf #1}, #2}
\def \PASP #1 #2 {{\em Publ. Astr. Soc. Pacific\/} {\bf #1}, #2}
\def \NAT #1 #2 {{\em Nature\/} {\bf #1}, #2}
\def\etal{{\it et al.}\ }
\def\Msun{{~\rm M}_\odot}
\def\EE#1{\times 10^{#1}}
\def\lsim{\!\!\!\phantom{\le}\smash{\buildrel{}\over
  {\lower2.5dd\hbox{$\buildrel{\lower2dd\hbox{$\displaystyle<$}}\over
                               \sim$}}}\,\,}

\documentstyle{memsait}
\input epsf.sty
\begin{opening}
\title{MODELING SUPERNOVA EMISSION AT LATE TIMES} % ALL CAPITAL LETTERS PLEASE !!!
\author{Cecilia Kozma}
\institute{Stockholm Observatory, SE-133 36 Saltsj\"{o}baden, Sweden\\
}
\date{} % DO NOT INSERT ANY DATE HERE !!!
\end{opening}

\begin{document}

%\oddpagefooter{\sf Mem. S.A.It., Vol. ??, ??}{}{\thepage}
%\evenpagefooter{\thepage}{}{\sf Mem. S.A.It., Vol. ??, ??}
\oddpagefooter{}{}{} % LEAVE AS IT IS !
\evenpagefooter{}{}{} % LEAVE AS IT IS !
\ 
\bigskip

\begin{abstract}
We compare model calculations with observations of supernovae at late times
to infer the time evolution of temperature, ionization and line emission. 
Here we mainly report on our results from our modeling of SN 1987A.
We discuss the oxygen mass from the modeling of line fluxes.
Line profiles 
show the distribution of the elements and the importance of including 
time dependence in the calculations.
We discuss different approaches to determine the $^{44}$Ti-mass.
\end{abstract}

\section{Introduction}
As the supernova ejecta expand they become optically thin in the
continuum after $\sim$ 200 days, and  it is possible to directly 
probe the interior of the ejecta. The main purpose of our 
modeling of the spectra and line fluxes 
is to put constraints on  the nucleosynthesis taking place 
in the progenitor, as well as during the explosion itself.
See Kozma \& Fransson (1998b) for a detailed discussion.

There are several observational indications that mixing occurred 
in the explosion of SN 1987A.
One is the early emergence of X-rays (Dotani \etal 1987;
Sunyaev \etal 1987) and $\gamma$-rays (Matz \etal 1988).
The effects of mixing can also be seen in the observed line profiles
at late times (Stathakis \etal 1991; Spyromilio, Stathakis, \& Meurer 1993;
Hanuschik \etal 1993). By modeling line profiles we are able to study the
distribution of mass  of different elements.
This mass distribution gives us information on the 
hydrodynamics taking place in the explosion
(e.g., M\"uller, Fryxell, \& Arnett 1991; Herant \& Benz 1992; Herant, Benz, \& Colgate 1992). 
In modeling the line profiles we can also see the importance of including
all the different composition regions, as well as time dependence in our
calculations. 

The modeled emission gives information on the energy source 
powering the ejecta.
Possible energy sources at late times are radioactive isotopes, 
a central compact object, and circumstellar interaction.
The radioactive isotopes formed in the
explosion, and which power the ejecta at subsequent
times are $^{56}$Ni, $^{57}$Ni, and $^{44}$Ti. 
After $\sim$ 1700 days $^{44}$Ti is the dominant isotope.
The formation of this isotope is sensitive to the occurrence
of $\alpha$-rich freeze-out (e.g., Woosley \& Hoffman 1991; 
Timmes \etal 1996).
The amount of this element in the ejecta therefore directly probes 
the  explosion mechanism itself. 

Although our model is quite general and may be applied to any supernova, 
as long as it is not dominated by
circumstellar contribution, we will here concentrate
on the results from our study of SN 1987A.

\section{Model}
Our modeling of SN 1987A is described in detail in Kozma \& Fransson (1998a). 
Here we will
just make a short summary.
The explosion model we use here for our abundances is the
10H model (Woosley \& Weaver 1986; Woosley 1988).
Our density and velocity structure is based on hydrodynamical calculations
(e.g., Herant \& Benz 1992) and on line profiles (Phillips \etal 1990;
Meikle \etal 1993). 
Our treatment of the thermalization of the
non-thermal $\gamma$-rays and positrons from radioactive decays is 
described in Kozma \&
Fransson (1992), and is based on the Spencer-Fano equation (Spencer \& Fano 1954). 
The radioactive isotopes we include are $^{56}$Co, $^{57}$Co, and $^{44}$Ti.

We solve the thermal and ionization balances time dependently, as well as
the level populations of the most important ions. 
The total number of transitions included in our calculations are $\sim$ 6400.

We take dust 
absorption into account by assuming optically thick clumps with a 
constant covering 
factor of 0.40 (Lucy \etal 1991; Wooden \etal 1993) after 600 days, and  
a linear increase from 0 to 0.40 in the time interval 350 to 600 days.

\section{Uncertainties}

One of the uncertainties in our modeling is our treatment of line blanketing.
The Sobolev approximation is used for the line transfer which in general is 
a good  approximation for the high velocity expansions in supernovae.
However, it does not take into account any interaction between different lines
or different regions in the ejecta. Especially in the UV, where there
are a lot of resonance lines, such interaction is expected to be important.
The importance of UV-scattering decreases with time as the matter thins.
The blocking of the UV-lines affects the emergent spectrum by degrading
the UV-photons to photons of lower energies. It also changes the UV-field 
within the ejecta which affects the ionization balance for mainly 
ions with low ionization potentials and low abundances, such as Na I,
 Mg I, Si I, Ca I, Fe I.

Another aspect of supernova evolution, which we do not include, is the 
formation of molecules. In SN 1987A CO and SiO were observed already 112 
days after the explosion (Oliva, Morwood, \& Danziger 1987; Spyromilio \etal 1988). 
Modeling of CO formation in SN 1987A has been done by e.g. Gearhart, Wheeler, 
\& Swartz (1999), who find a CO-mass of $\sim$ $10^{-4} \Msun$ at 200 days.
As molecules can be strong coolers, they can 
greatly influence the temperature at their formation sites. 
In order to form CO both carbon and oxygen should be abundant. The most 
favorable site for this to happen is therefore in the oxygen-carbon rich 
regions. 
The most favorable place for SiO to form is in the transition region
between the silicon-rich and oxygen-rich regions. 
The regions suitable
for molecule formation have little mass, which explains the low mass of 
molecules in the ejecta.

The importance of the H$_2$-molecule has been discussed by 
Culhane \& McCray (1995). They calculate the abundance of  H$_2$
theoretically as it has not been observed. They also discuss
how the resonance scattering in the Lyman and Werner bands 
may affect the emergent UV-spectra.

There are several observational indications of dust formation in the
ejecta 
after $\sim$ 400 - 500 days (Roche \etal 1989; Lucy \etal 1991; Meikle \etal 1993).
Dust absorbs much of the harder radiation, thermalizes it,
and reradiates it in the IR. The dust absorption affects both the line
fluxes and profiles.
We treat dust absorption in a simplified way in our 
models, as mentioned above. 
Our treatment of the dust assumes it to be located in optically 
thick clumps. This is a reasonable assumption as 
the effects of dust absorption are seen in both 
optical and IR lines (Lucy \etal 1991), consistent with optically
thick clumps. 

However, we do not include dust cooling in our calculations..
When we compare our modeled spectra (Figure 2) and light curves 
(Figure 4) with observations we find a general good agreement. We 
therefore believe that our temperature structure 
well describes the conditions in the ejecta,
and that dust cooling is not significant in the line
emitting regions.
 
In the decay of $^{44}$Ti $\rightarrow$ $^{44}$Sc + $e^+$ $\rightarrow$ $^{44}$Ca + $\gamma$
the positrons dominate the non-thermal energy input at late times. In our model we
assume complete trapping of these positrons in the iron-rich regions, containing
the newly synthesized iron. Chugai \etal (1997) find that the intensity of 
the Fe II lines can 
be explained only if trapping of the positrons from ${}^{44}$Ti is
efficient in the iron-rich parts of the ejecta. 
Again based on the fact that we find such  good agreement 
between modeled and observed fluxes (Figures 2 and 4), assuming full 
trapping of positrons, we believe this to be
a good assumption. A leakage of positrons would result in a much faster decline
of the light curves than observed.
 
\section{Temperature and ionization evolution}

The temperatures in the different composition regions 
in the supernova ejecta evolve differently,
depending on composition and density as 
discussed in detail in Kozma \& Fransson (1998a). 

The temperature in the 
inner, heavy-element regions decreases slowly with time until it reaches a temperature
of $\sim$ 2000 K. At that time far-IR lines replace optical lines as 
the dominant source of cooling and the temperature suddenly drops to
$\lsim 500$ K. The reason for this is a thermal instability
setting in. The cooling by far-IR lines is 
insensitive to the temperature at temperature much larger then $\sim$ 500 K. 
This sudden drop in temperature is referred to as the IR-catastrophe 
(Axelrod 1980; Fransson \& Chevalier 1989; Fransson 1994).

The time for the onset of the IR-catastrophe varies for the 
different compositions.
For example, in the iron-rich region 
the drop in temperature  
sets in at $\sim$ 600
days (Figure 1). In Figure 1 one can also see the gradual transition from 
optical to near-IR, and to far-IR lines as the IR-catastrophe sets in.

In the hydrogen- and helium-rich regions this IR-catastrophe is not seen. Here
adiabatic cooling is more important, and the temperature decreases more 
slowly, as $T \propto t^{-2}$.

\begin{figure}
\epsfysize=6cm % fix the y-dimension and scales x-dim. to y-dim.
\epsfxsize=12cm % fix the x-dimension and scales y-dim. to x-dim.
% Feel free to do the choice you prefer but do not exceed the x-dimension
% of the text lines
\hspace{3.5cm}\epsfbox{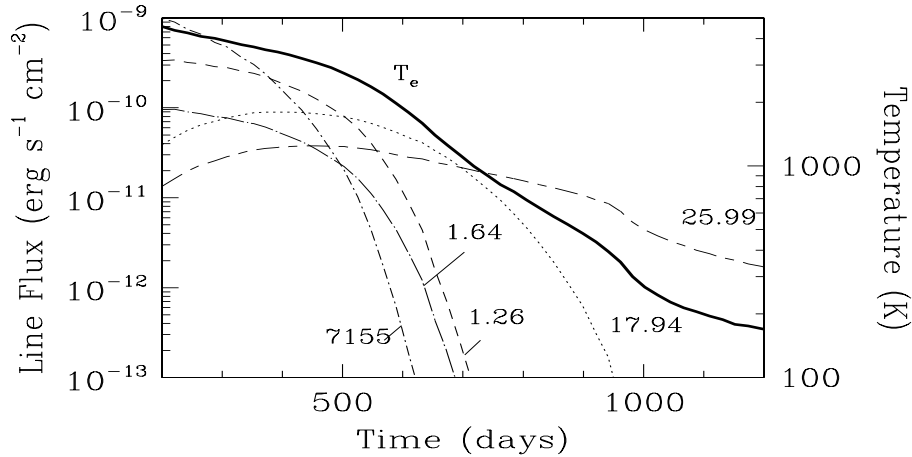} %for centering: act on hspace argument 
\caption[h]{Evolution of temperature for the iron-rich region 
in SN 1987A, together with the fluxes of the strongest Fe II lines. 
The IR-catastrophe is seen to set in at 
$\sim$ 600 days.}
\end{figure}

In our modeling we find it important to include time dependence.
After $\sim$ 800 - 900 days the recombination and cooling time scales 
become longer 
than the radioactive decay time scales and the steady state assumption
is no longer valid. This freeze-out effect is discussed in Fransson \&
Kozma (1993), and is crucial for modeling the bolometric light curve,
as well as 
individual line fluxes and profiles.

\section{Line fluxes}

\begin{figure}
\epsfysize=6cm % fix the y-dimension and scales x-dim. to y-dim.
\epsfxsize=12cm % fix the x-dimension and scales y-dim. to x-dim.
% Feel free to do the choice you prefer but do not exceed the x-dimension
% of the text lines
\hspace{3.5cm}\epsfbox{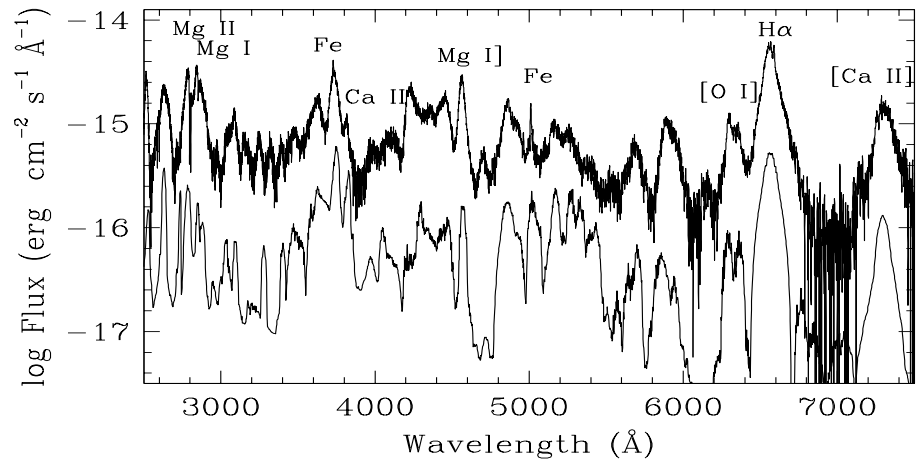} %for centering: act on hspace argument 
\caption[h]{A comparison between a modeled (lower) and observed (upper) 
spectrum of SN 1987A at 2870 days (priv. comm. SINS collaboration, Kirshner \etal).}
\end{figure}

Here we will discuss a couple of interesting points concerning our 
modeling of line fluxes and spectra. A thorough  discussion of the line 
emission is given in Kozma \& Fransson (1998b).
In Figure 2 we compare a preliminary calculated spectra with observations 
and find quite good agreement. In this calculation we have included line
scattering in a simplified way. We find that in particular the 
[Ca II] $\lambda\lambda$7291,7324 lines and the IR-triplet are very
sensitive to the assumption of line scattering. Radiative pumping of the 
calcium H, K lines by UV emission, mainly from Fe I, is very important 
at this time.

Because of its importance for the oxygen mass determination we now 
discuss the [O I] $\lambda\lambda$6300,6364 lines in some detail.
In these lines we can see the temperature
evolution of the oxygen-rich regions (dotted line in Figure 3). Up to $\sim$
800 days these lines are formed by thermal excitation to the $^1D$ level. 
At that time the IR-catastrophe sets in and the temperature drops rapidly.
At later epochs the temperature is too low in these regions 
for thermal excitation to be
of any importance, and non-thermal excitation is responsible for
the line emission. 
We are able to model the thermal part of the light curve quite well.
For the non-thermal part, however, we run into 
problems. There are HST observations of the 
[O I] $\lambda\lambda$6300,6364 lines up to
day 3597, and we underproduce the line fluxes up to factor of almost 10. 
In Kozma \& Fransson (1998b) we discuss in detail our modeling of these lines.
We include contributions from the triplet levels to the $^1D$ level, which 
are sensitive to the composition in the oxygen rich regions. We discuss the
importance of photoionization, we try different filling factors, 
etc. Even if we push all our assumptions as far as we can in order
to maximize the  [O I] $\lambda\lambda$6300,6364 fluxes we are still
not able to reproduce the observations. 

\begin{figure}
\epsfysize=6cm % fix the y-dimension and scales x-dim. to y-dim.
\epsfxsize=12cm % fix the x-dimension and scales y-dim. to x-dim.
% Feel free to do the choice you prefer but do not exceed the x-dimension
% of the text lines
\hspace{3.5cm}\epsfbox{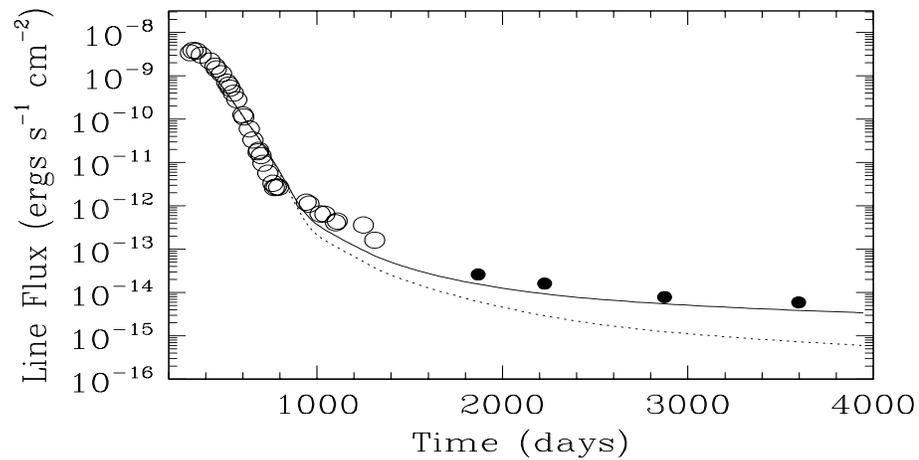} %for centering: act on hspace argument 
\caption[h]{The flux of the 6300 \AA-feature for SN 1987A. The dotted line shows the flux solely  
due to [O I] $\lambda\lambda$6300,6364. The solid line is the sum of the [O I] and the [Fe I] fluxes at 6300 \AA~. 
The open dots are data from Danziger \etal (1991), 
and the filled dots are HST data (priv. comm. SINS collaboration, Kirshner \etal).}
\end{figure}

Because of this failure, we are now happy to report that 
the solution to this problem is to be found in blending of the 
[O I] $\lambda\lambda$6300,6364 lines with a fairly strong [Fe I] 
multiplet (Fransson, Kozma \& Wang 1999). 
While this is unimportant for the thermal part, the low non-thermal flux 
is dominated by the [Fe I] lines.
In Figure 3 we show a preliminary 
calculation of the 6300 \AA-feature, including both the [O I] and [Fe I]
emission (solid line), together with observations. The dotted line in
Figure 3 shows the contribution only from the 
[O I] $\lambda\lambda$6300,6364 lines. 
This agreement therefore makes our determination of the oxygen mass in 
SN 1987A considerably more firm. 
In Kozma \& Fransson (1998b) we find a value of $\sim$ 1.9 $\Msun$ 
of oxygen enriched gas.

\section{Line profiles}

The line profiles provide a
tool to probe the distribution of the different elements in the ejecta.
In Kozma \& Fransson (1998b) we compare our modeled line profiles of H$\alpha$, He I 2.058 $\mu$m,
[O I] $\lambda\lambda$6300,6364 with observations and estimate the 
mass and distribution of hydrogen, helium and oxygen. 

In Fransson \etal (1999) we continue this work  and model 
the H$\alpha$ line profile for different epochs up to 4000 days. 
We find that the hydrogen envelope, as reflected in the line wings and 
extending from 2000 to 6000 km s$^{-1}$, becomes increasingly 
important with time. This is an effect of the freeze-out becoming 
more important in the outer, low density regions of the envelope.

Our modeled line profiles of iron, e.g., [Fe II] 17.94 $\mu$m and 25.99 $\mu$m 
also show line wings out to 6000 km s$^{-1}$ (Fransson \& Kozma 1999), 
showing that the primordial iron within the hydrogen envelope contributes
significantly, due to freeze-out. The contribution 
from primordial iron to the  [Fe II] 25.99 $\mu$m line flux can be
seen in Figure 5 (dash-dotted line).

\section{$^{44}$Ti-mass}

The energy input to the ejecta at late times is the decay of radioactive
isotopes formed in the explosion.
The three radioactive isotopes that 
subsequently dominate the energy input are 
$^{56}$Co, $^{57}$Co, and $^{44}$Ti. The decay of $^{56}$Co dominates up to
$\sim$ 800 - 900 days, thereafter $^{57}$Co becomes an increasingly 
important energy source. 
After $\sim$ 1700 days the energy input to the ejecta is dominated by
$\gamma$-rays and positrons from the decay of $^{44}$Ti. 

The amount of $^{56}$Co in the ejecta is accurately determined from 
modeling of the bolometric light curve and is found to be $\sim$ 0.07 $\Msun$
for SN 1987A. Also the amount of $^{57}$Co can be inferred from the
bolometric light curve if the effects of freeze-out are properly taken into
account (Fransson \& Kozma 1993). Other ways to estimate the  
$^{57}$Co-mass are from studies of [Fe II] and [Co II] IR-lines (Varani \etal 1990; Danziger \etal 1991) and observation of the 122 keV $^{57}$Co line 
(Kurfess \etal 1992). The inferred mass of $^{57}$Ni expelled is 
$\sim$ 0.0033 $\Msun$.

The $^{44}$Ti isotope is very interesting as it, together  with $^{57}$Co,
is formed in the very supernova explosion (e.g., Woosley \& Hoffman 1991), 
and the amount  
formed is sensitive to the conditions prevailing there.
There are several ways to try to estimate the mass of $^{44}$Ti
in the ejecta of SN 1987A. 
To model the bolometric light curve and compare it 
with observations is no longer a fruitful approach.
Observational uncertainties of the bolometric light curve are large 
for times later than $\sim$ 1200 days, as more and more of the emission 
emerges in the IR. For example, Bouchet \etal (1996) find that at 2172 days
 $\sim$ 97 \% of the energy is emitted in the IR, out of which only 
 $\sim$ 1 \% can be directly observed. 

A  better approach is to observe and model the different broad bands.
We then directly avoid the observational uncertainties.
In Kozma \& Fransson (1999) we discuss our modeling of broad band 
photometry and by comparing the models with observations we 
reach a preliminary estimate of the   
$^{44}$Ti-mass of $(1.5 \pm 1.0)\EE{-4} \Msun$.
In Figure 4 the light curves for the B and V bands are shown. For the
three models we use a $^{44}$Ti-mass of  $0.5\EE{-4}, 
1.0\EE{-4}$, and $2.0\EE{-4} \Msun$, respectively.

\begin{figure}
\epsfysize=6cm % fix the y-dimension and scales x-dim. to y-dim.
\epsfxsize=12cm % fix the x-dimension and scales y-dim. to x-dim.
% Feel free to do the choice you prefer but do not exceed the x-dimension
% of the text lines
\hspace{3.5cm}\epsfbox{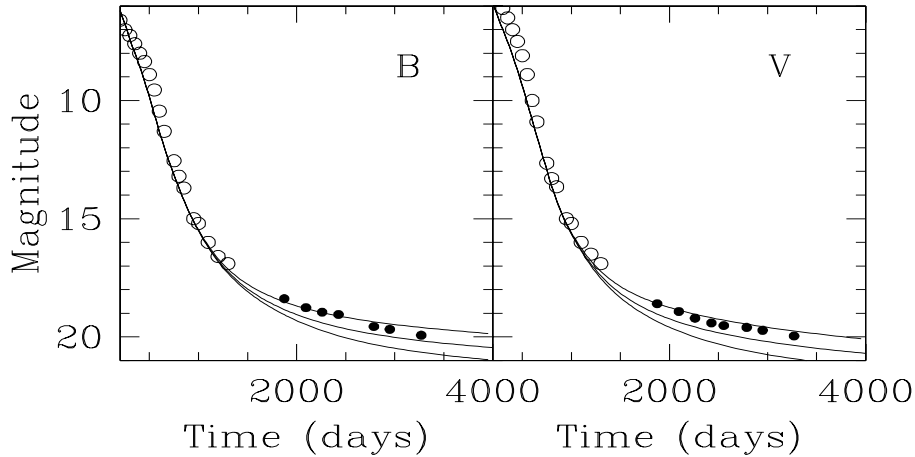} %for centering: act on hspace argument 
\caption[h]{The light curves for the B and V bands up to 4000 days 
for SN 1987A.
The three models contain $0.5\EE{-4}, 1.0\EE{-4}, 2.0\EE{-4} \Msun$ of 
$^{44}$Ti, respectively. The observations are from Suntzeff \& Bouchet (1990) and Suntzeff \etal (1991) (open dots), and HST data (Suntzeff 1999) 
(filled dots).}
\end{figure}

Still another approach to estimate the mass of $^{44}$Ti is to look at individual
line fluxes. The line used for such an estimate has to be carefully chosen.
For example, the strong and well-understood H$\alpha$ line turns out to
be a bad choice. The  H$\alpha$ emission has large
contributions  from the envelope regions at later times where the freeze-out
effect dominates (Kozma \& Fransson 1999; Fransson \etal 1999). The  H$\alpha$ emission is therefore   
very insensitive to the instantaneous energy input. 

\begin{figure}
\epsfysize=6cm % fix the y-dimension and scales x-dim. to y-dim.
\epsfxsize=12cm % fix the x-dimension and scales y-dim. to x-dim.
% Feel free to do the choice you prefer but do not exceed the x-dimension
% of the text lines
\hspace{3.5cm}\epsfbox{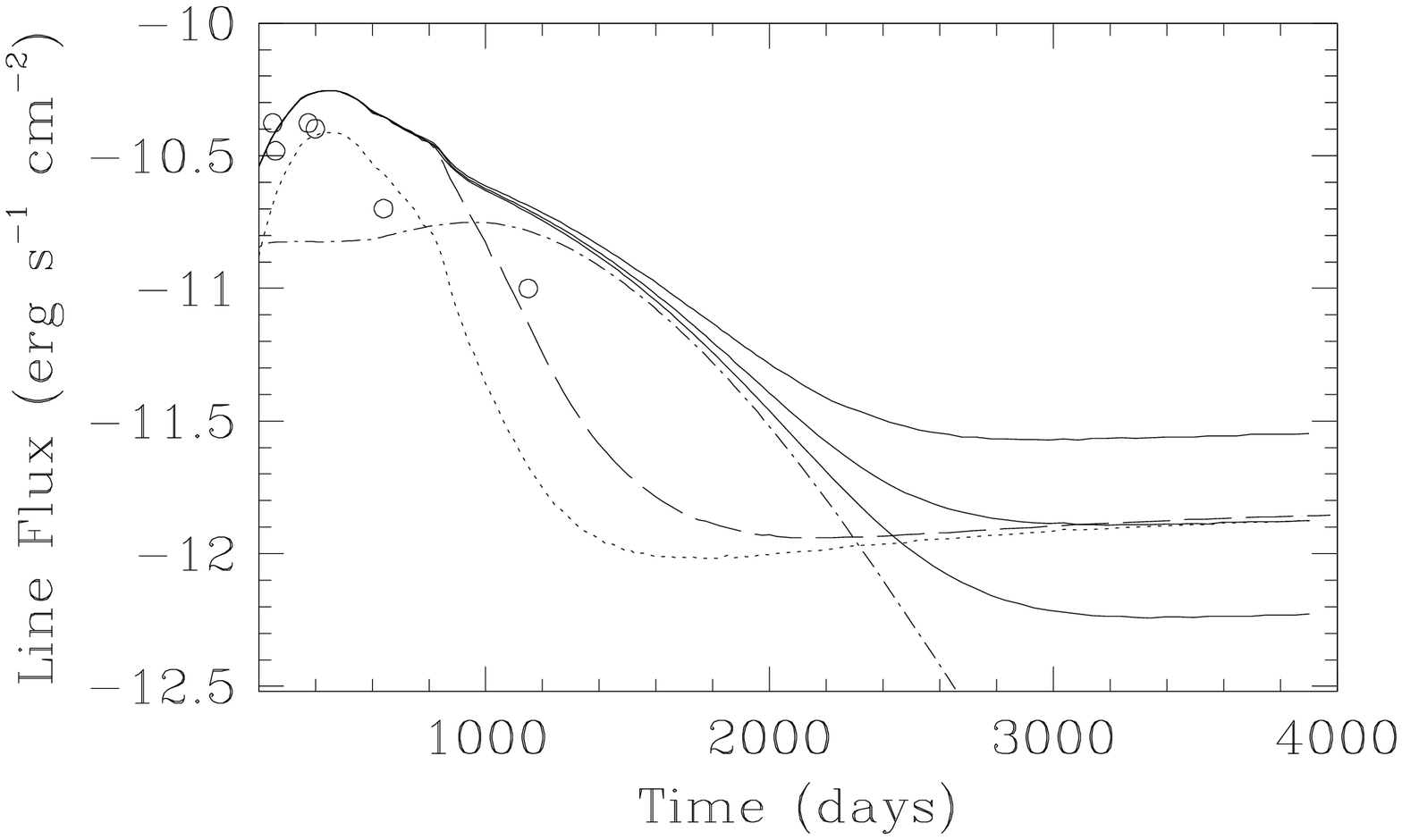} %for centering: act on hspace argument 
\caption[h]{The flux of the [Fe II] 25.99 $\mu$m line for SN 1987A. 
The three solid lines show the line fluxes for [Fe II] 25.99 $\mu$m
for three masses of $^{44}$Ti; $0.5\EE{-4}$, $1.0\EE{-4}$, and $2.0\EE{-4}$ $\Msun$.
The dashed line is a model assuming steady state and M($^{44}$Ti)=$1.0\EE{-4}$ $\Msun$. The dotted line represents the flux contribution from 
newly synthesized iron in the iron-rich regions, while the dash-dotted line
is the contribution from the primordial iron in the hydrogen-rich regions.} 
\end{figure}

A better choice would be the iron lines. At these late times they originate 
in the iron-rich region within the core where freeze-out is much less 
important. There are however problems also with these lines. Many of the
iron lines in the optical and IR are heavily blended, which makes it
difficult to infer fluxes of individual lines. Another problem is the
uncertainty of the 
atomic data, in addition to incomplete atomic models.
The best individual line is [Fe II] 25.99 $\mu$m. This 
arises from collisional excitations and the uncertainties in the
atomic data related to the recombination cascade is avoided.
Figure 5 shows the calculated flux for the [Fe II] 25.99 $\mu$m line
for SN 1987A. The three  $^{44}$Ti-masses used are $0.5\EE{-4}$, $1.0\EE{-4}$, 
and $2.0\EE{-4}$ $\Msun$ (the three solid lines). 
As shown in this figure the
flux is almost proportional to the $^{44}$Ti-mass after $\sim$ 2500 days.
Also shown in Figure 5 is a steady state model (the dashed line), for the 
M($^{44}$Ti)=$1.0\EE{-4}$ $\Msun$ case. Steady state is a good approximation
(for this line)
for times earlier than $\sim$ 1000 days, and times later than $\sim$ 3000 days.
For times in between, time dependence has to be included in order to model
the line fluxes accurately. The reason for this can be understood from  
 the origin of the iron flux. The dotted line in Figure 5 shows the 
contribution to the total line flux from the newly synthesized iron, in the
iron-rich regions. The dash-dotted line, on the other hand, shows the 
contribution from the primordial iron within the hydrogen-rich regions.
Between $\sim$ 1000 and $\sim$ 2400 days the primordial iron 
dominates the line flux. But even at earlier times the primordial 
iron contributes significantly to the flux. 
This iron mostly resides in the low density hydrogen envelope which is
most sensitive to the freeze-out effect, and therefore 
to the assumption
of time-dependence. After $\sim$ 3000 days the newly synthesized iron
dominates the emission. The iron-rich regions is much less affected by freeze-out
and the emission directly reflects the energy input. 
Therefore this line is a good tracer of the $^{44}$Ti-mass.
We find in our modeling that  approximately half of 
the supernova's emission at 4000 days emerges in this line.
Therefore the [Fe II] 25.99 $\mu$m line flux 
is one of the most reliable ways to determine the $^{44}$Ti-mass. 

ISO observations of the [Fe II] 25.99 $\mu$m are reported 
in Lundqvist \etal (1999), and give 
an upper limit on the observed [Fe II] 25.99 $\mu$m
flux on day 3999. These observations together with model calculations
give an estimate of the  upper limit of the  $^{44}$Ti-mass of 
$\sim 1.4\EE{-4} \Msun$.
A third, and future way, to estimate M($^{44}$Ti) is to observe the 
$\gamma$-ray line at 1.156 MeV with instruments like INTEGRAL (Leising 1994).

\acknowledgements
I would like to thank Claes Fransson and Peter Lundqvist 
for stimulating 
discussions and for useful comments on the manuscript.
I also thank the SINS team (P.I. R.~Kirshner) for using data prior  
to publication.
This research was supported by the Swedish Natural Science Research
Council, the Swedish National Space Board and 
the Knut and Alice Wallenberg Foundation.

% References. We avoided using the \bibitem commmand since we found it is
% somewhat platform-dependent. We also avoided using the \cite{keyword}
% command since we found it cumbersome. However, if you are an expert 
% LateX user you may use the various LateX tools for the references 
% provided they give the same printout formats of the examples given here.

\end{document}